# Acoustically regulated optical emission dynamics from quantum dot-like emission centers in GaN/InGaN nanowire heterostructures

S. Lazić[1], E. Chernysheva[1], A. Hernández-Mínguez[2], P. V. Santos[2] and H.P. van der Meulen[1]

[1]Departamento de Física de Materiales, Instituto "Nicolás Cabrera" and Instituto de Física de Materia Condensada (IFIMAC), Universidad Autónoma de Madrid, 28049 Madrid, Spain
[2]Paul-Drude-Institut für Festkörperelektronik, Hausvogteiplatz 5-7, 10117 Berlin, Germany

E-mail:  lazic.snezana@uam.es

**Abstract.** We report on experimental studies of the effects induced by surface acoustic waves on the optical emission dynamics of GaN/InGaN nanowire quantum dots. We employ stroboscopic optical excitation with either time-integrated or time-resolved photoluminescence detection. In the absence of the acoustic wave, the emission spectra reveal signatures originated from the recombination of neutral exciton and biexciton confined in the probed nanowire quantum dot. When the nanowire is perturbed by the propagating acoustic wave, the embedded quantum dot is periodically strained and its excitonic transitions are modulated by the acousto-mechanical coupling. Depending on the recombination lifetime of the involved optical transitions, we can resolve acoustically driven radiative processes over time scales defined by the acoustic cycle. At high acoustic amplitudes, we also observe distortions in the transmitted acoustic waveform, which are reflected in the time-dependent spectral response of our sensor quantum dot. In addition, the correlated intensity oscillations observed during temporal decay of the exciton and biexciton emission suggest an effect of the acoustic piezoelectric fields on the quantum dot charge population. The present results are relevant for the dynamic spectral and temporal control of photon emission in III-nitride semiconductor heterostructures.

## 1. Introduction

Compared to conventional planar semiconductor devices, the implementation of semiconductor heterostructures on a nanowire platform provides a fundamental building block for nanoscale photonic and optoelectronic devices. For example, by embedding optically active quantum dots (QDs) into individual nanowires (NWs) scalable miniaturized architectures to produce non-classical light states (i.e. single photons and entangled photon pairs) can be achieved [1-3]. In addition, the NWs can be easily detached from the as-grown sample and transferred onto a foreign substrate, which facilitates their integration into modern on-chip technologies. However, to fully exploit their potential, appropriate techniques are needed to probe and to dynamically control the fundamental physical effects on a nanoscale. In the last years, there has been an increasing interest in the use of radio-frequency (rf) surface acoustic waves (SAWs) for the in-situ dynamic control and manipulation of the emission characteristics from small-scale semiconductor heterostructures, including NW QDs. These concepts were built on schemes that have been previously established for planar semiconductor systems. Compared to all-electrical approaches, which typically require selective doping and sophisticated nanofabricated electrical contacts on individual NWs with sub-micrometer dimensions, the SAW-spectroscopy has proven to be a powerful contactless technique for: the control of acousto-electrically induced conveyance of charge carriers and dissociated excitons across the NW [4-7], dynamic programing of NW QD occupancy state [8], precisely timed carrier injection into and extraction from NW QDs for low-jitter single photon emission [6,7], tuning of the NW QD radiative optical transitions by the oscillating strain and piezoelectric SAW fields [9, 10], as well as coherent control of NW-based nanophotonic resonators [11].

Because the SAWs propagate at the speed of sound, their wavelengths in semiconductor heterostructures are typically in the micrometer and sub-micrometer range, thus covering acoustic frequencies from several tens of megahertz up to the gigahertz range. Moreover, SAWs can propagate almost without dissipation over macroscopic distances exceeding several millimeters. In this way, many nanostructures located within the acoustic pathway can simultaneously interact with the strain and piezoelectric fields accompanying the propagating SAW. In addition, individual nanostructures being perturbed by the SAW can be independently identified and studied using all-optical techniques.

Most reports on SAW-modulated nanowire QD structures to date are limited to III-arsenide material systems. Applying SAW-techniques to group III-nitrides is vital, as their wide bandgap, large exciton binding energy and band offsets make them an ideal candidate for high-power and high-temperature device applications. Also, compared to III-arsenides, the high sound velocities and the stronger electromechanical coupling coefficients in III-nitrides allow for high frequency applications [12]. So far, SAW-experiments carried out on individual III-nitride NWs have been mainly focused on electric transport experiments [13] and no detailed and complete study on the SAW-governed optical emission dynamics has been reported in these systems.

In our recent work, performed on InGaN QD-like emission centers embedded in GaN NWs mechanically transferred on a lithium niobate (LiNbO$_3$) SAW chip, we have demonstrated the SAW-induced tuning of the QD energy levels and its excitonic binding energies [14]. Our previous study have also shown that these NW QDs emit almost fully linearly polarized single photons up to ~80 K [15], whose output can be clocked at the acoustic frequency [7]. In this contribution, we review our experiments on the effect of SAW-induced strain and piezoelectric fields on the photon emission dynamics from these dot-in-a-nanowire heterostructures. We employ time-integrated and time-resolved micro-photoluminescence (μ-PL) spectroscopy under stroboscopic optical excitation with a variable phase shift between phase-locked pump laser pulses and the propagating acoustic fields. Our results show that, when the nanowire is subjected to a propagating acoustic wave, the embedded quantum dot is periodically stretched and compressed and its excitonic transitions are modulated by this oscillating strain field. The observed spectral shifts directly reflect dynamic changes in the NW QD electronic energy levels. Depending on the recombination lifetime of the involved excitonic transitions, we can resolve acoustically driven radiative processes and assess their dynamic spectral changes during the acoustic cycle. At high acoustic amplitudes, we observe waveform distortions in the NW QD spectral response resulting from the propagation of harmonic waves along the LiNbO$_3$ substrate. In addition, we show that all characteristic signatures observed in the temporal decay of the QD excitonic emission can be readily understood by the underlying acousto-electrically induced charge carrier dynamics, as confirmed by numerical calculations of the SAW-induced oscillating piezoelectric field. Compared to previously studied QD systems, this work, therefore, reports on the following novel aspects: *(i)* stroboscopic and full time-domain acousto-optoelectric spectroscopy in III-nitride QD systems, *(ii)* detection of higher-order harmonic sound waves on LiNbO$_3$ surface using nanowire-based sensor QDs and *(iii)* SAW-regulated simultaneous ambipolar injection of photo-generated electrons and holes into a single quantum emitter. These results are relevant for the dynamic control and manipulation of single photon emission in III-nitride semiconductor heterostructures.

## 2. Methods
*2.1. Growth of dot-in-a-nanowire heterostructures*
The GaN nanowires hosting InGaN QDs were fabricated via a site-controlled "bottom-up" approach on GaN-on-sapphire templates by RF-plasma-assisted molecular beam epitaxy (MBE). The NW site control was achieved using a titanium mask-based selective area growth (SAG) technique. More information on the sample fabrication and employed growth parameters can be found elsewhere [2,16,17]. The NWs have typical dimensions of ~500 nm in length and ~200 nm in diameter and exhibit a hexagonal cross section with "pencil-like" top morphology. A thin InGaN section with varying thickness and indium compositions (up to 30 nm and ~20%, respectively, depending on the growth facet) was embedded in the NW apex and capped with a ~50 nm thick GaN [2,17]. As detailed in Refs. [7,14,15,16], the QD-like emission centers under study are formed by alloy composition fluctuations in the topmost part of the InGaN region.

*2.2. Acoustic delay line*
A schematic diagram of the sample configuration is depicted in figure 1(a). All optical experiments were performed on NWs transferred onto a 128° Y-cut LiNbO$_3$ substrate photo-lithographically patterned with an acoustic delay line consisting of two interdigital transducers (IDTs). The IDTs are used for all-electrical excitation and detection of SAWs. For SAW excitation, an rf signal of the appropriate frequency is applied to one of the IDTs. During the SAW propagation along the delay line at the sound velocity of $v_{SAW}$~3980 m/s, as illustrated in figure 1(b), the strain field generated through the inverse piezoelectric effect is accompanied by an electric field induced by the direct piezoelectric effect. When the wave reaches the opposite IDT it is transformed back into an rf signal which can be detected and analyzed.

The use of a highly piezoelectric LiNbO$_3$ crystal provides strong strain and electric fields of the propagating SAW. Because the acoustic power is confined to within approximately one SAW wavelength (both inside and outside) from the surface of the substrate, the SAW fields induced in LiNbO$_3$ crystal extend to the optically active NW heterostructures deposited on the surface and can modulate their electronic band structure.

The IDTs employed in this study are floating electrode unidirectional transducers (FEUDT) [18] designed to facilitate excitation of multiple harmonics of a fundamental wavelength $\lambda_{SAW,1}$=35 μm. The experiments reported here were carried out using the third harmonic with wavelength of $\lambda_{SAW,3}$=11.67 μm at room-temperature, which corresponds to a resonant frequency and period of $f_{SAW}$~332 MHz and $T_{SAW}$~3 ns, respectively. Details on the rf characteristics of the IDTs and the SAW transmission line (in particular, the scattering parameters $S_{11}$ (reflection) and $S_{21}$ (transmission and insertion loss) around the resonance) are given in Ref. [7].

*2.3. Micro-photoluminescence experiments*
μ-PL spectroscopy was employed to assess the emission characteristics of the NW QDs as well as their SAW-induced recombination dynamics. For the low-temperature μ-PL measurements the SAW-chip containing dispersed NWs was mounted on a cold finger cryostat equipped with an optical access window and coaxial connections for the application of rf signals to the IDTs. The measurements were carried out on a single NW with its growth axis almost perpendicular to the LiNbO$_3$ surface and the SAW propagation direction [7,14] (cf. scanning electron micrograph (SEM) in figure 1(a)). Electron-hole pairs were photo-generated in the InGaN NW apex using the $\lambda_{exc}$=442 nm excitation wavelength of either a continuous wave (cw) helium cadmium (HeCd) laser or a pulsed diode laser emitting $\tau_{laser}$ <

70 ps long pulses. The experiments were performed under direct optical excitation with the laser beam focused using a 100× microscope objective onto ~1.5 μm diameter spot at the targeted NW position. The NW photoluminescence was collected via the same objective and dispersed using a single grating monochromator with spectral resolution of ~350 μeV. The signal is detected either time-integrated by a liquid $N_2$ cooled multichannel Si-CCD camera or time-resolved using a Si avalanche photodetector (APD, PDM series from PicoQuant) of ~50 ps temporal resolution, combined with the time-correlated single photon counting (TCSPC, PicoHarp 300 from PicoQuant) electronics.

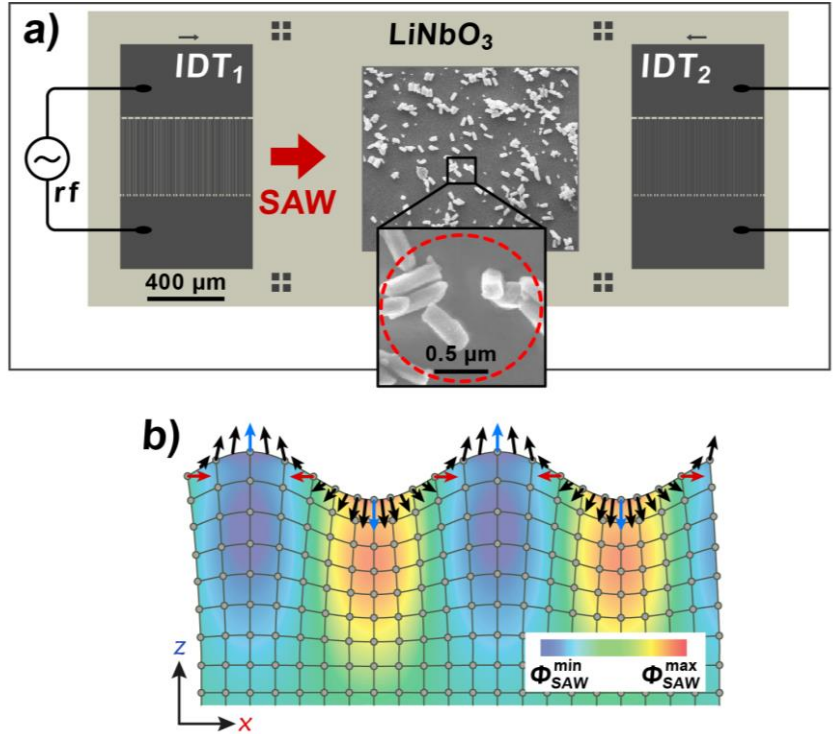

**Figure 1.** (a) Layout of the SAW-chip formed by a pair of interdigital transducers (IDTs) patterned on the surface of a $LiNbO_3$ crystal. The GaN/InGaN nanowire (NW) heterostructures are mechanically dispersed in between the two IDTs. Inset: Magnified SEM image of dispersed NWs. Dashed red circle outlines the optically probed NWs. (b) Calculated mechanical vibration (mesh), associated piezoelectric potential $\Phi_{SAW}$ (color coded) and electric fields (arrows) of a propagating sinusoidal acoustic wave.

To monitor the acousto-mechanic and acousto-electric effects on the probed NW QD emission dynamics we employ a stroboscopic μ-PL technique [19,20] with either time-integrated or time-resolved detection. For stroboscopic optical excitation, the laser trigger was derived from the rf generator and its repetition frequency ($f_{laser}$) was phase-locked to the SAW by setting $f_{laser}=f_{SAW}/8$. By electrically delaying the laser trigger in steps of 200 ps (corresponding to ~6.5% of the SAW period) relative to a fixed SAW phase, the carriers can be photo-excited at any well-defined time during the acoustic cycle. For time-resolved stroboscopic detection, the laser reference (sync) signal and the electrical output of the APD were directly connected to the inputs of the TCSPC module. In this way, the TCSPC electronics was started on a laser trigger and stopped on a photon detection by the APD. The center and the width of the spectral detection window for time-resolved experiments were determined by the spectrometer's diffraction grating position and the width of its slits. For all time-resolved experiments, the reference time t=0 was calibrated by measuring the reflectance of the laser pulses, which are used to excite the PL.

## 3. Results and discussions
### 3.1. Acoustically induced spectral tuning

Figures 2(a) and 2(b) present time-integrated µ-PL spectra recorded on the CCD camera from a single NW QD reported in Refs. [7,14] at a temperature of T≈10 K under cw optical excitation. In the spectrum measured with no SAW applied we identify signatures arising from recombination of excitonic complexes corresponding to the neutral exciton (X=1e+1h) at $E_{0,X}$=2.5643 eV and biexciton (XX=2e+2h) at $E_{0,XX}$=2.5871 eV [7,14]. As the $IDT_1$ is driven at its resonance frequency of ~337.5 MHz, a large fraction of the applied rf power is transformed to the SAW. Note that this resonance is slightly shifted to higher frequencies compared to its room-temperature value due to the increased SAW phase velocity at low temperatures. As this SAW couples to the NWs, the embedded QD is dynamically strained and its sharp X and XX emission lines are modulated by the deformation potential coupling [14]. Averaged over time, this modulation gives rise to an apparent splitting of the X and XX emission energies. The measured X (XX) acousto-mechanical response in figure 2(a) (2(b)) was quantified by fitting (red dashed curves) the experimental data (black solid curves) using [21,22]:

$$I(E) = I_0 + \frac{1}{T_{SAW}} \frac{2A}{\pi} \int_0^{T_{SAW}} \frac{w}{4 \cdot (E - (E_0 + \Delta E \cdot \sin(2\pi \cdot f_{SAW} \cdot t)))^2 + w^2} dt \qquad (1)$$

which corresponds to a time-averaged PL spectrum of Lorentzian peak with intensity $A$, width $w$ and its center emission energy $E_0$ sinusoidally tuned in time with amplitude $\Delta E$. Due to the time-integrated detection and the unlocked excitation scheme (i.e. cw excitation), the recorded data provides an averaged picture of the SAW-governed carrier recombination. In this way, we can assess the full bandwidth of the SAW-driven periodic spectral tuning of the QD optical transitions, which for the highest SAW intensity at $P_{rf}$=+25 dBm reaches a maximum value of $\Delta E_X$=1.51 meV for X and $\Delta E_{XX}$=1.62 meV for XX [14].

By actively locking $f_{laser}$ to the RF signal used to excite the SAW, we record in figure 2(c) (2(d)) phase-resolved µ-PL spectra of the X (XX) emission line. In these experiments, the detection was averaged over all SAW phases, while the pulsed laser excitation took place at two different times during the acoustic cycle shifted by $T_{SAW}/2$. The X emission (cf. figure 2(c)) shows the same SAW-split PL line for both excitation phases. On the contrary, for the XX peak (cf. figure 2(d)), the two PL traces are phase-shifted by 180°. To better understand these results, we present in figures 2(e) (2(f)) normalized PL spectra for the X (XX) acquired under the same stroboscopic detection conditions as in figure 2(c) (2(d)) and plotted in false color representation as a function of the photon energy (vertical axis) and stroboscopic phase delay $\Delta\varphi$ between the laser pulses and the SAW excitation (horizontal axis). By tuning $\Delta\varphi$ (cf. Section 2.3) we can optically pump the NW QD at any phase of the SAW cycle and, thus, obtain full phase information of the SAW-governed emission dynamics. All spectra were taken from the same NW QD over two acoustic cycles at $P_{rf}$=+16 dBm applied to the input transducer $IDT_1$. As we tune $\Delta\varphi$, the XX peak exhibits clear and pronounced sinusoidal spectral oscillations with the SAW period, while the X shows a rather broad emission independent of the SAW phase at which the optical excitation takes place.

An acoustically induced periodic modulation of the emission energies similar to the one of the XX transition was reported for QDs [8,9,23,24] and nanocavities [20] in the III-As material system. It was attributed to the deformation potential coupling induced by dynamic SAW strain field. In the same way, we can argue that the spectral changes observed in figure 2 are dominated by the strain contribution. In fact, as discussed in our previous work in Ref. [14], for both X and XX radiative transitions, the maximum (minimum) emission energies occur at SAW phases corresponding to the maximum compression (tension). For the spatial orientation of the NW determined by SEM imaging (cf. figure 1(a)), only the transverse component (of the order of few kV/cm) of the oscillating SAW-induced piezoelectric field is present at these particular SAW phases (blue arrows in figure 1(b)). Thus, the piezoelectric contribution to the observed spectral changes via Stark effect is negligible compared to the strong built-in field (~1 MV/cm) along the NW c-axis and cannot be resolved in our experimental data.

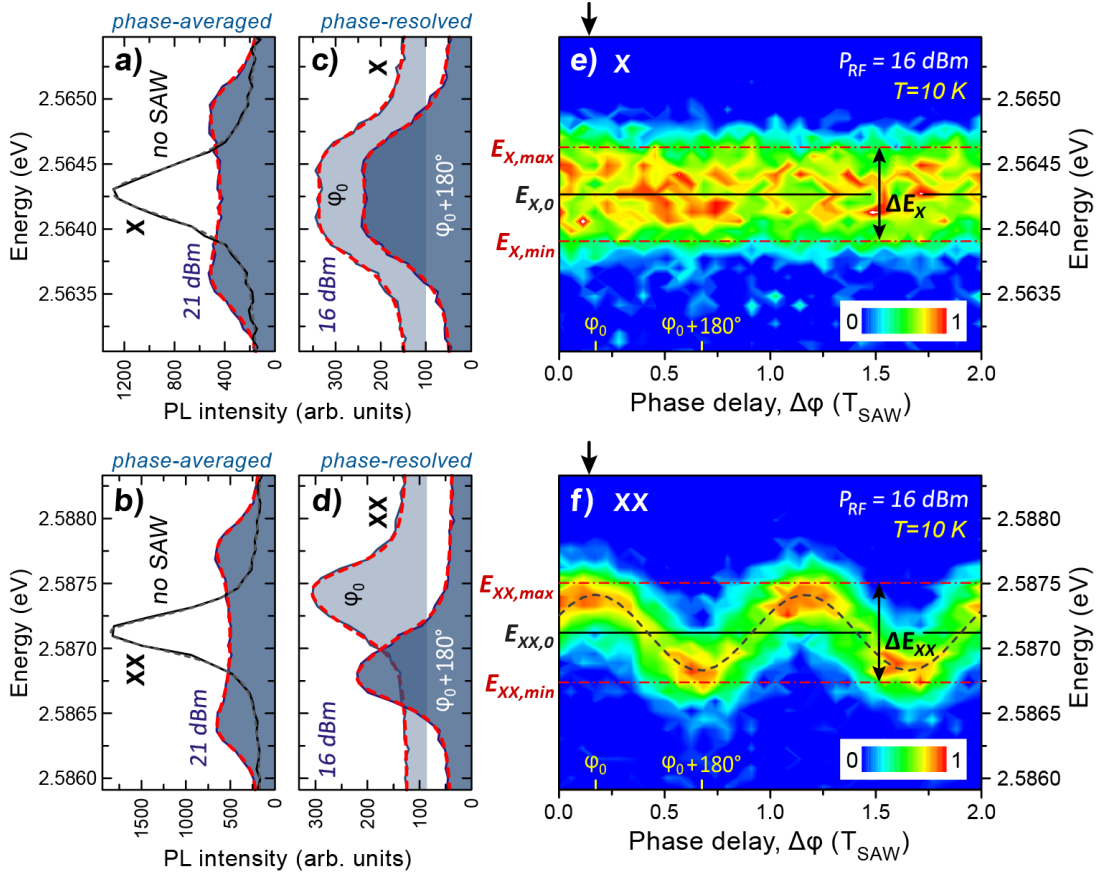

**Figure 2.** (a-d) Time-integrated μ-PL of a single NW QD without (unshaded traces) and with SAWs of different amplitudes (shaded traces) applied under (a),(b) cw and (c),(d) stroboscopic optical pulsed excitation. The rf power applied to $IDT_1$ (cf. Figure 1) is indicated for each panel. The stroboscopic optical excitation phases for the two PL traces plotted (offset with respect to each other for clarity) in (c) and (d) differ by 180°. They correspond to local SAW phases at which the probed NW QD is under maximum compressive and tensile strain (marked as $\varphi_0$ and $\varphi_0+180°$ in (e) and (f), respectively). Red dashed curves in (a)-(d) are fits of the experimental data to equation (1) and grey dashed curves in (a)-(b) are fits to an unperturbed Lorentzian peak. (e),(f) Intensity maps of the time-integrated (e) X and (f) XX normalized PL intensities as a function of the phase delay Δφ (horizontal axis, in units of the acoustic period $T_{SAW}$) and photon emission energy (vertical axis) for $P_{rf}$=+16 dBm. The black solid lines mark the energies of unperturbed (i.e. without a SAW) X ($E_{0,X}$) and XX ($E_{0,XX}$) transitions. The dash-dot red lines represent their values at SAW phases corresponding to the maximum compressive ($E_{X(XX),max}$) and tensile ($E_{X(XX),min}$) strain. The dashed black line is a guide to the eye highlighting the SAW-driven XX spectral modulation.

As demonstrated in [10,19], stroboscopic optical excitation combined with time-averaged detection can only resolve periodically driven processes for optical transitions with time decay constant of $\tau_{PL}$ <$T_{SAW}/2$. To explain the findings of figures 2(e) and 2(f), we plot in figure 3(a), the time dependence of the X and XX PL intensities after pulsed excitation in the absence of the SAW. These measurements were done with the spectrometer exit slit width set to register the FWHM bandwidth of the measured optical transition, as determined from the unperturbed X and XX PL peaks in figure 2(a) and 2(b), respectively. Both PL traces show exponential decays with time constants of $\tau_X=2.78\pm0.09$ ns for the X and $\tau_{XX}=0.78\pm0.07$ ns for the XX transition. These values suggest that the SAW-driven periodic changes in the photon emission energy can only be seen for a fast radiative decay of the XX transition, whose lifetime is four times shorter than the SAW period (i.e. $\tau_{XX}\approx T_{SAW}/4$). On the contrary, owing to its long lifetime, which is approximately equal to the duration of a SAW cycle ($\tau_X\approx T_{SAW}$), the X emission is averaged over the full bandwidth of the SAW-induced spectral tuning presented in figure 2(a). As a consequence, no SAW-driven sinusoidal spectral oscillations can be observed for the X emission line.

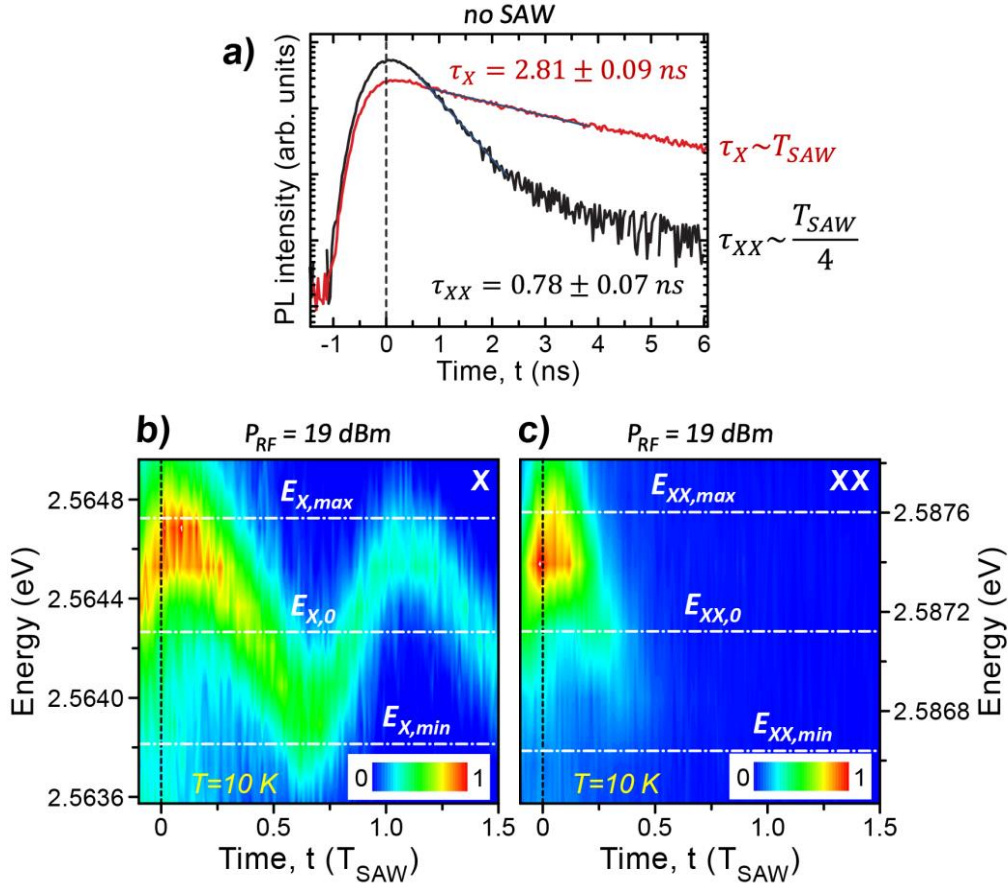

**Figure 3.** (a) PL decay curves of the spectrally filtered exciton (X) and biexciton (XX) with no SAW applied. The estimated decay time constant is comparable to the acoustic period of $T_{SAW}=2.96$ ns for the X or equal to approximately $T_{SAW}/4$ for the XX. (b),(c) False color plot of the time-resolved (horizontal axis, given in units of $T_{SAW}$) X (b) and XX (c) normalized PL intensity as a function of the emission energy (vertical axis) for $P_{rf}=+19$ dBm.

To get a deeper insight into the underlying dynamics of excitonic transitions in the acoustically modulated QD confinement potential, we have registered the time evolution of the PL at the APD detector under SAW-synchronized photo-excitation and time- and spectrally resolved detection conditions. The carriers are optically excited at a precisely defined phase of the acoustic cycle ($\varphi_0$) marked with a vertical black arrow in figure 2(e),(f). The wavelength scan was carried out by moving the monochromator diffraction grating in steps corresponding to 0.04 nm while acquiring the time-resolved PL traces for each grating position. The spectral bandwidth of the signal sent to the APD was set to ~20% of the SAW-split X (XX) peak, as estimated from time-averaged PL spectra similar to those in figure 2(a) (2(b)) recorded on the CCD under cw optical pumping for a given rf power ($P_{rf}$=+19 dBm). The PL time-transients measured from both X and XX are presented in figures 3(b) and 3(c), respectively. As discussed previously, when the SAW interacts with the optically probed NW QD, the energy of its excitonic transitions is constantly changing during the SAW period. Contrary to time-integrated experiments of figure 2(c-f), we now see that for a slow X recombination, we can clearly resolve the time-dependent SAW-driven sinusoidal spectral oscillations during the decay of its emission intensity. The same is, however, not observed for the XX decay in figure 3(c) because of its short lifetime, which is shorter than $T_{SAW}/2$. This is in accordance with the results shown in figures 2(d) and 2(f) since, for both time-averaged and time-resolved detection, the effective collection time is limited to the XX lifetime.

*3.2. RF power-dependent waveform distortions*
For high powers $P_{rf}$ applied to $IDT_1$, the NW QD spectral response to the propagating SAW becomes distorted from the sinusoidal waveform detected at lower rf powers (cf. figure 2(f)). An example is given in figure 4(a) for $P_{rf}$=+24 dBm. In order to determine the origin of the signal deformation, we compare in figure 4(b) the rf-waveform applied to $IDT_1$ (grey points) with the one arising from the SAW transmitted across the delay line and detected by $IDT_2$ (red points). These measurements were performed at room-temperature using an oscilloscope on a SAW delay line like the one used in the optical spectroscopy experiments, but without dispersed NWs. At large powers of the rf-amplifier employed in our experimental setup, the rf-excitation waveform applied to $IDT_1$ slightly deviates from an ideal sinusoidal shape (black dotted dashed curve) due to the nonlinear electrical behavior of the rf-amplifier at voltage levels close to its maximum power output. The waveform detected on $IDT_2$, however, exhibits a strongly distorted shape. Although changes in the SAW waveform can be induced by the presence of unwanted harmonic frequencies sent to $IDT_1$ by the rf-amplifier (its frequency domain equivalent recorded by a spectrum analyzer confirmed the presence of harmonic signal components of differing relative amplitudes (not shown)), we attribute the observed distortions mainly to the generation of harmonics of the driving acoustic frequency due to nonlinear acoustic interactions in the $LiNbO_3$ substrate at high rf power excitations [25,26]. The resulting harmonic-frequency waves can then effectively interact to reconstruct the fundamental mode, but the extent to which this occurs depends on the strength of nonlinear dispersion effects, which scales with the SAW amplitude. These nonlinear phenomena can, thus, lead to the distortion of an initially sinusoidal acoustic waveform [27]. In the present studies, the nonlinear effects can already take place within the FEUDT transducer because its multi-harmonic capability, determined by the finger geometry and metallization ratio [28], is compatible with the harmonics of the driving acoustic frequency generated by the nonlinear interactions. Note, in particular, that these nonlinear effects are responsible for the electrical detection of the distorted shape on $IDT_2$ (red points in figure 4(b)). This signal is proportional to the electric potential $\Phi_{SAW}(t)$ induced by the piezo-mechanical coupling. The black dashed curve superimposed on this signal in Fig. 4(b) is a fit to the experimental data using Fourier-series analysis [24], thereby revealing the presence of higher-order harmonic components.

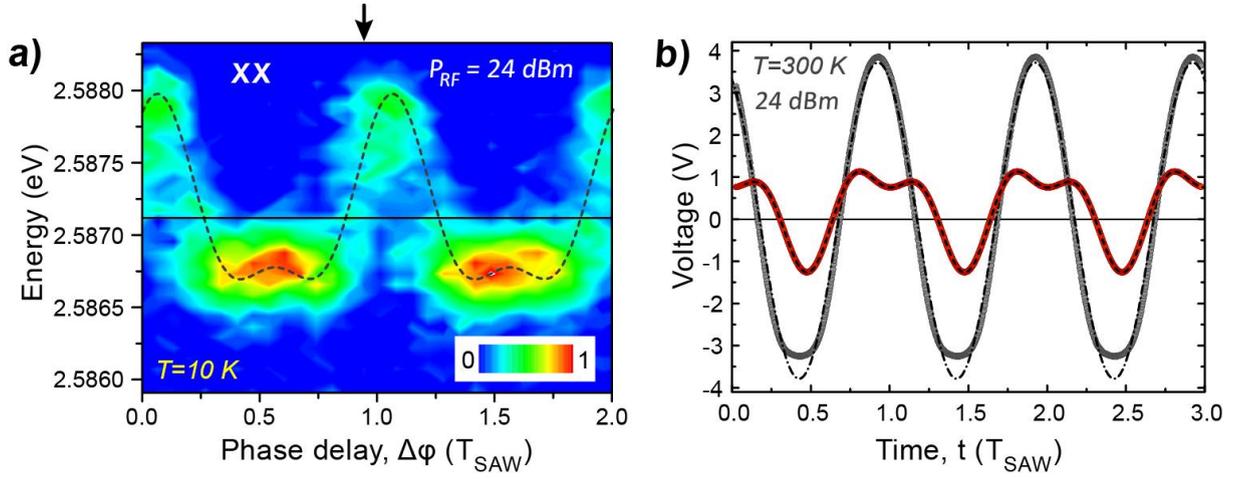

**Figure 4.** (a) Color coded plot of the time-integrated XX PL intensity as a function of phase delay $\Delta\varphi$ (in units of $T_{SAW}$, horizontal axis) and photon emission energy (vertical axis) for $P_{rf}$=+24 dBm. The data are normalized and plotted on the same color scale as in figure 2(f). The black solid line marks the XX transition energy in the absence of a SAW while the dashed black line is a guide to the eye outlining the SAW-driven XX spectral modulation. (b) Room-temperature oscilloscope traces of the rf signal applied to sending transducer $IDT_1$ (grey points) and detected on receiving transducer $IDT_2$ (red points) together with their corresponding best fits to an ideal sine-wave (dash-dot curve) and a four-term Fourier-series in a sine-wave basis (dashed curve), respectively.

In contrast, the time-dependent SAW-governed spectral modulation of the excitonic transitions in our NW QD is dominated by the deformation potential coupling [14], which is proportional to the spatial derivative of the mechanical displacement of the Rayleigh-type SAW and, thus, of the associated piezoelectric potential $\Phi_{SAW}$ [24]. The resulting changes inflicted by the aforementioned nonlinear phenomena on the waveform of the acoustic signal transmitted across the $LiNbO_3$ substrate are, therefore, reflected in the spectral response of our sensor QD deposited on the substrate's surface. In fact, the acoustically controlled modulation of its XX emission wavelength in figure 4(a) is well reproduced (with the out-of-phase relation) by the SAW signal electrically detected on $IDT_2$ of the $LiNbO_3$ SAW-chip with no dispersed NWs (cf. figure 4(b)).

*3.3. Acoustically mediated ambipolar charge injection*

Lower panels in figures 5(a) and 5(b) present the temporal decay of the X and XX PL at moderate ($P_{rf}$=+16 dBm) and high ($P_{rf}$=+24 dBm) acoustic powers, respectively. In both cases, the spectral detection window was chosen to collect the entire SAW-induced spectral modulation bandwidth of the measured excitonic line. We employ weak optical pumping for which the densities of charge carriers optically excited in the InGaN section of the NW apex are low and do not significantly screen the SAW-induced piezoelectric fields. For each $P_{rf}$ value the carriers are photo-excited at the precise phase of the corresponding SAW-induced spectral modulation marked by a vertical black arrow in figures 2(e),(f) and 4(a), respectively. As the TCSPC electronics was triggered using the same signal derived from the rf generator (cf. Section 2.3), the t=0 zero-time reference in the PL time-transients of Fig. 5 corresponds to the time during the acoustic cycle at which the pulsed photo-excitation takes place.

The measured time-resolved PL profiles reveal the dependence of the SAW-mediated QD emission dynamics as a function of the local phase of the SAW. Contrary to the unperturbed PL transients in figure 3(a) as well as the acoustically mediated full time-domain stroboscopy of the excitonic transitions in figures 3(b) and 3(c), the stroboscopic time-resolved measurements displayed in lower panels of figures 5(a) and 5(b) show weak changes in the X and XX PL intensity during their respective decays. For the (slow) X decay at moderate SAW amplitudes (cf. figure 5(a)), the PL intensity oscillations appear twice in the SAW period ($T_{SAW}=1/f_{SAW}\approx2.96$ ns) yielding multiple characteristic signatures in the X PL time-transient. At high acoustic powers (cf. figure 5(b)), these oscillations become more pronounced and their temporal sequence changes. In the case of the (fast) XX recombination, depending on the stroboscopic excitation phase, at most one characteristic signature in the emission intensity can be resolved during its decay (cf. peak at ~$0.3T_{SAW}$) for each laser excitation cycle ($T_{laser}=1/f_{laser}\approx23.7$ ns). Its temporal occurrence coincides with that of the X, as expected for the X and XX emission originating from the same QD. Moreover, by changing the excitation phase by $\Delta\varphi'$, the times at which these oscillations appear shift by $\Delta t= \Delta\varphi'\cdot T_{SAW}/2\pi$ (not shown).

To understand these experimental observations, we show in upper panels of figure 5(a) and 5(b) the calculated time dependence of the longitudinal $F_X$ (i.e. along SAW propagation) and transverse $F_Z$ (i.e. perpendicular to the substrate surface) components of the SAW-induced electric field for $P_{rf}$=+16 dBm and $P_{rf}$=+24 dBm, respectively. The numerical modelling of the interactions between our $In_XGa_{1-x}N$ QD (with x≈20% [17]) and the propagating SAW strain and piezoelectric fields was performed following the procedure described in Ref. [24]. This was done assuming the ground state optical transition under SAW-induced deformation potential perturbation (i.e. without taking into account the excitonic effects). The acoustic frequency used in these calculations is the actual experimental value obtained from the low-temperature (T=10 K) IDT response monitored by a vector network analyzer (not shown). By comparing the observed temporal emission characteristics of the QD to the simulated time evolution of the two electric field components, we find that for all $P_{rf}$ values the increase of X and XX emission intensities takes place at the acoustic phases (marked with horizontal red arrows in figures 1(b) and 5(a-c)) corresponding to the maximum and minimum values of the longitudinal electric field.

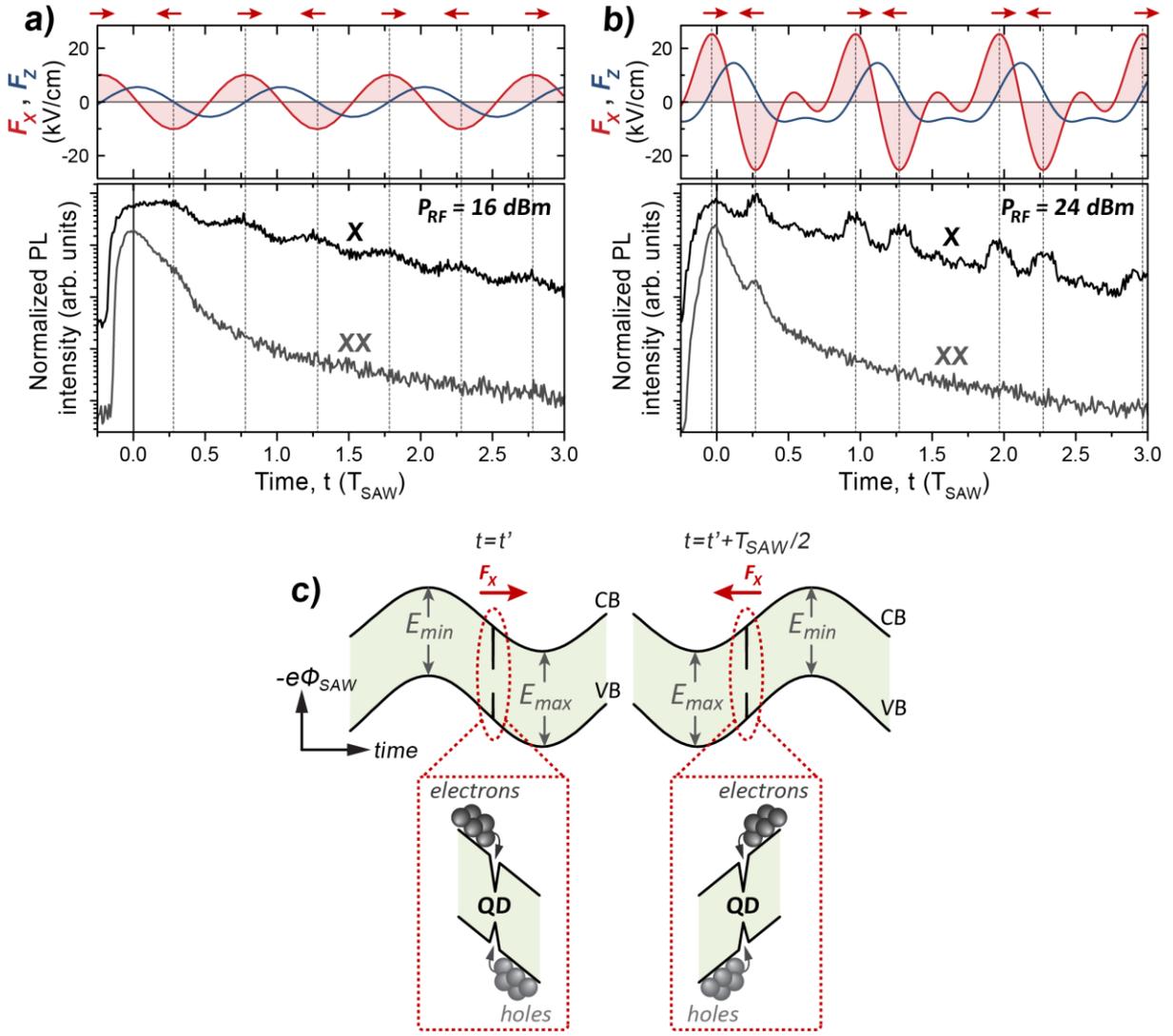

**Figure 5.** (a),(b) Upper panels: Time dependence of the longitudinal $F_X$ (red shaded curves) and transverse $F_Z$ (blue curves) electric field components at the position of the probed NW QD. Lower panels: Time-resolved traces of the SAW-driven X and XX µ-PL for (a) $P_{rf}$=+16 dBm and (b) $P_{rf}$=+24 dBm applied to $IDT_1$. The traces are vertically offset for clarity. (c) Upper panels: Schematics of the conduction (CB) and valence (VB) band edge modulation induced by the SAW strain and piezoelectric fields within the GaN/InGaN NW heterostructure. Lower panels: Potential energy and charge dynamics across the NW QD at SAW phases corresponding to maximum and minimum $F_X$ value. The QD embedded in the surrounding InGaN region acts as an optically active trap for electrons and holes.

In addition, the amplitude of SAW-mediated emission intensity oscillations scales with the strength of the piezoelectric field. For low acoustic powers, the piezoelectric field components are weaker and the aforementioned changes in the PL intensity can no longer be resolved (not shown). As $P_{rf}$ increases, the waveform distortions discussed in section 3.2 set in and the $F_X$ and $F_Z$ time evolution during the acoustic cycle is changed (cf. figure 5(b)). As in the case of lower $P_{rf}$ values in figure 5(a), the resulting sequential increase of the emission intensity during its decay again reflects the SAW phases at which the amplitude of $|F_X|$ attains its maximum value.

Because the diameter of the laser spot is larger than the probed NW heterostructure hosting the QD and its energy is above the absorption edge of the InGaN section in the NW apex, the carriers are generated in the entire topmost InGaN region [7,14,15]. Since the studied QD is, in fact, a potential trap induced by indium content fluctuations within the InGaN apex [14,15], it acts as an optically active deep trap for electrons and holes photo-excited in the InGaN region (cf. figure 5(c)). The negative binding energy of the involved XX transition (~ -22.9 meV) is also indicative of strong QD confinement [14]. We, therefore, relate the observed PL intensity oscillations to the photo-generation in a continuum of states of the InGaN region surrounding the QD, where the longitudinal electric field of the SAW induces temporal carrier dynamics. This, in turn, leads to an acoustically regulated simultaneous transfer of photo-excited electrons and holes into the energetically lower QD states. The correlated PL intensity oscillations between X and XX transitions in the experimental data of figures 5(a) and 5(b) also corroborate the assertion of SAW-controlled charge injection into the QD. As illustrated in figure 5(c), the charge transfer to the QD takes place at SAW phases corresponding to the maximum magnitude of the SAW-induced longitudinal electric field $F_X$, irrespective of its orientation (i.e. along or against the SAW propagation direction). As a consequence, the recombination probability peaks at the same points during the SAW cycle as the $|F_X|$ amplitude. The transverse electric field component $F_Z$, on the other hand, has negligible effect on the charge population of the SAW-modulated QD potential since its contribution is small compared to the large built-in field acting in the direction of the NW axis (cf. section 3.1). The experimentally observed simultaneous injection of the two carrier species (i.e. electrons and holes) into the QD confined energy levels differs quite significantly from the previous findings reported in other QD systems, in which the SAW-regulated spatio-temporal carrier dynamics leads to preferential sequential carrier injection [5-8,10,29,30] or extraction [9].

## 4. Summary

We have investigated acoustically driven dynamics of the optical emission from InGaN QDs embedded in epitaxially grown GaN/InGaN nanowire heterostructures mechanically transferred to a LiNbO$_3$ SAW-chip. We found that the excitonic transitions in these QDs are sensitive to the propagating strain and piezoelectric acoustic fields. Using a stroboscopic µ-PL technique based on both time-averaged and time-resolved detection with variable phase shift between phase-locked SAW and pulse optical excitation we can monitor acoustically driven recombination processes over time scales defined by the SAW. Depending on the PL lifetime of different excitonic complexes, we can obtain full phase information of the SAW-driven emission dynamics. For high SAW amplitudes, we observe distortions from the sinusoidal waveform in the NW QD spectral response, which we attributed to the propagation of harmonic-frequency waves along the LiNbO$_3$ substrate. Moreover, the time dependence of the SAW-controlled PL emission exhibits characteristic signatures, which are manifested in the sequential increase of the emission intensity. These PL intensity oscillations observed during the decay of both exciton and biexciton optical transitions stem from acoustically regulated transfer into the QD of charge carriers photo-generated in the surrounding InGaN region. This charge transfer and capture into the QD's confined energy states is assisted by the SAW longitudinal piezoelectric field, as confirmed by comparing our experimental data to numerical simulations of oscillating SAW-induced electric fields.

**Acknowledgements**
The authors thank J.M. Calleja form the Universidad Autónoma de Madrid for scientific discussions & Ž. Gačević and E. Calleja from ISOM-DIE at the Universidad Politécnica de Madrid for the epitaxial growth of nanowire heterostructures. E.C. (S.L.) acknowledges the Spanish MINECO FPI (RyC-2011-09528) grant. The project was partially funded by the Spanish MINECO Grants MAT2014-53119-C2-1-R and MAT2017-83722-R.